\RequirePackage{fix-cm}
\documentclass[smallextended]{svjour3}       
\usepackage{bm}
\usepackage{empheq}
\usepackage{epsfig}
\usepackage{graphicx}
\usepackage{epstopdf}
\usepackage{mathtools}
\usepackage{amssymb}
\usepackage{amsmath}

\smartqed  

\def\be{\begin{equation}}
\def\ee{\end{equation}}
\def\bea{\begin{eqnarray}}
\def\eea{\end{eqnarray}}
\def\beaN{\begin{eqnarray*}}
\def\eeaN{\end{eqnarray*}}
\def\ed{\end{document}}
\def\bit{\begin{itemize}}
\def\eit{\end{itemize}}

\def\Del{\Delta}
\def\del{\delta}

\def\k{\kappa}
\def\alf{\alpha}

\def\di{\partial}

\def\half{{\textstyle{1 \over 2}}}
\def\~{\tilde}

\def\m{\label}
\def\l{\left}
\def\r{\right}

\def\goto{\rightarrow}

\def\const{\rm const}

\newcommand{\goh}{\mathfrak{h}}

\usepackage{graphicx}
\begin{document}

\title{A point mass and continuous collapse to a point mass in general relativity
} \subtitle{}

\titlerunning{A point mass and continuous collapse to a point mass in general relativity}

\author{Alexander N. Petrov
}


\institute{A. N. Petrov \at
              Moscow MV Lomonosov State University, Sternberg Astronomical
              Institute,  Universitetskii pr. 13, Moscow 119992, Russia \\
              Tel.: +7(495)7315222\\
              Fax: +7(495)9328841\\
              \email{alex.petrov55@gmail.com}           
           }

\date{Received: date / Accepted: date}

\maketitle

\begin{abstract}
An original way of presentation of the Schwarzschild black hole in the form of a point-like  mass with making the use of the Dirac $\delta$-function, including a description of a continuous collapse to such a point mass, is given. A maximally generalized description restricted by physically reasonable requirements is developed. A so-called field-theoretical formulation of general relativity, being equivalent to the standard geometrical  presentation of general relativity, is used. All of the dynamical fields, including the gravitational field, are considered as propagating in a  background (curved or flat) spacetime. Namely these properties allow us to present a non-contradictive picture of the point mass description. The results can be useful for studying the structure of the  black hole true singularities and could be developed for practical calculations in models with black holes.

\end{abstract}

\section{Introduction: motivation and goals}
 \m{Introduction}

 The Schwarzschild solution \cite{Landau_Lifshitz_1975,Misner_Thorn_Wheeler_1973} in general relativity (GR) has been obtained under the simplest assumptions: it has to be a vacuum, spherically symmetric and static one. A choice of the integration constant, $M$, corresponds to the assumption that the solution is induced by isolated gravitational masses with the Newtonian asymptotic behaviour. All of these means that a source could be a point particle (point mass). However, one finds that the Schwarzschild solution is a black hole solution, whereas the notion of a point mass in GR becomes unclear. Indeed, attempting to reach a point localized at the center one meets, earlier and later, the horizon of events and another structure of a black hole.

 In spite of that, it is quite interesting to represent a description of a black hole in the form of a point particle. The reasons are:

First, it is a fundamental interest to find out a point-like derivation. It could be achieved with making the use of an additional mathematical technique or equivalent formalism, which do not contradict to predictions of GR.

Second, Newtonian gravity is the optimistic example for describing a point mass. The Newtonian equation is
 \be
 \nabla^2 \phi = -4\pi\rho(r),
 \m{a-1_a}
 \ee
 where  $\nabla^2$ is the Laplace operator, $\phi$ is the gravitational potential and $\rho(r)$ is the mass density. To describe a point particle one assumes that the mass distribution is presented in the form, $\rho = M\delta(\bm r)$, where Dirac's $\delta$-function satisfies
 the Poisson equation,
 \be
 \nabla^2 \left({1\over r}\right) = -4\pi\delta(\bm r),
 \m{a-1}
 \ee
and the gravitational potential is presented by the Newtonian potential, $\phi = M/r$. As a result, one finds that the equation (\ref{a-1_a}), is satisfied in the whole space including even the point $r = 0$. Besides, the volume integration of $\rho$ over the whole space, the same as the surface integrating the left hand side of (\ref{a-1_a}), gives the accepted result for the total mass, $M$.

Third, many vacuum solutions in GR include singularities in curvature. Among them there are the Schwarzschild and Kerr black hole solutions, where the singularities are under the horizon of events. Usually, analyzing such solutions, one leads to a convention that the Einstein equations do not hold at such singularities. Nevertheless, many authors suggest a description of black holes, when the singularity is presented with making the use of Dirac's $\delta$-function. Then it could be interpreted as a matter source of a curved geometry in the Einstein equations. Many studies are developed in this direction, see, for example, \cite{Balasin_Nachbagauer_1993,Pantoja_Rago_2002,Heinzle_Steinbauer_2002} and references there in.

Fourth, in 2016, perhaps one of the greatest discoveries of all time has been stated. It is a direct detection by laser interferometer gravitational-wave observatories, LIGO and Virgo, of the gravitational waves produced in the events of coalescence of binary black holes \cite{GW150914,GW151226,GW-A-LIGO}. To be convinced in the prediction of the discovery a systematic theoretical study of dynamic compact relativistic objects was carried out and is carried out. As a rule, at an {\em initial} step the black holes are modeled by {\em point-like} particles presented by Dirac's $\delta$-function \cite{Damour_Jaranowski_Schafer_2001,Schafer_2003,Schafer_2011,Blanchet_2014}.

The programm of (\ref{a-1_a}) and (\ref{a-1}) in the Newtonian case, applied in order to describe a black hole in GR as a point mass, meets principal difficulties \cite{Narlikar_1985}. Let us show this. Substitute the black hole metric coefficients into the left hand side of the Einstein equations:
 \be
  G_{\mu\nu} = \kappa T_{\mu\nu}.
\m{a-3}
 \ee
 The Schwarzschild solution in the Schwarzschild coordinates is
 \be
 ds^2 = -\l(1-\frac{r_g}{r}\r)c^2dt^2 + \l(1-\frac{r_g}{r}\r)^{-1}dr^2 + r^2d\Omega^2,
\m{a-2}
 \ee
where, as usual, $d\Omega^2 = d\theta^2 + \sin^2\theta d\phi^2$, $r_g = 2MG/c^2$ and coordinates are numerated $ct = x^0$, $r = x^1$, $\theta = x^2$ and $\phi = x^3$. Assuming that the solution holds in the whole spacetime with the world line $r = 0$, the matter
distribution acquires the form \cite{Narlikar_1985}:
\bea
T_{0}{}^0 & =& T_{1}{}^1  = 0, \m{a-4}\\
T_{2}{}^2 & =& T_{3}{}^3   = - {{Mc^2} \over 2}~\delta({\bm r}).\m{a-5}
\eea

Zero's result (\ref{a-4}) shows that it is impossible to obtain the correct total mass in this case. The situation cannot be saved even if one remembers that the time coordinate and the radial coordinate change their sense inside the horizon \cite{Narlikar_1985}. Besides, a sense of $\delta$-function in (\ref{a-5}) under the horizon is lost because it is a function of time-like coordinate  Thus, the standard {\em geometrical} presentation of GR, where the true singularity is a spacelike singular hypersurface, is not so appropriate to represent the Schwarzschild solution as a point-like object.

Another reformulation of GR derived in the framework of so-called field-theoretical approach, see \cite{GPP_1984,Popova_Petrov_1988,Petrov_1993,Petrov_2008,Petrov+Co_2017}, can be an appropriate formalism to represent the Schwarzschild solution as a point mass.  The field-theoretical description is constructed with the help of a {\em simple} decomposition of the variables of the geometrical formulation of GR into a sum of
background and perturbed (dynamical) variables. All the dynamical fields, including the gravitational field,
are considered as a configuration of dynamical fields (field configuration) propagating in a  background (fixed, auxiliary) spacetime (curved or flat). By the construction, the geometrical and the field-theoretical formulations are equivalent, therefore any solutions to GR can be treated in the framework of both the approaches.

Up to now, in spite of significant efforts, see a review in section \ref{Preliminaries}, there is no an unique and complex strategy for describing point like objects and their formation in a non-contradictive way in GR. Therefore, the goal of the present paper is to close this gap, examining the Schwarzschild black hole solution. We represent its derivation in the form of a point-like object, including a continuous collapse to such a point-like mass and generalizing a description of such models by an appropriate way with physically reasonable restrictions. The mathematical basis of our study is the field-theoretical formulation of GR.

The paper is organized as follows. In Sect. \ref{Preliminaries}, we review works where the problem of a point particle in GR has been considered. Also we formulate necessary and reasonable requirements to achieve aforementioned goals of the paper. In Sect. \ref{sec:2}, we describe the main properties of the field-theoretical approach in GR which are used for constructing a point-like model and its formation in GR. We outline a general strategy for such constructions as well. In Sect. \ref{FC-delta}, we present the Schwarzschild solution as a point mass both in the Schwarzschild and Eddington-Finkelstein frames. After that this description is generalized being restricted by the physically reasonable requirements. In Sect. \ref{collapse}, we outline a continuous regular collapse to a point mass state modelled in Sect. \ref{FC-delta}. In Sect. \ref{discussion}, we discuss briefly the results.

\section{A short review and requirements for constructing a model}
 \m{Preliminaries}

Various approaches in gravity has been suggested to describe a point mass in GR a short review of which is given in this section.

In \cite{Balasin_Nachbagauer_1993},  a problem of vacuum geometries in GR with zero matter energy-momentum and singularities in curvature is analyzed. The authors advocate the use of a so-called distributional technique (regularization) to calculate geometrical quantities of manifolds equipped with a singular metric. What is interesting here, this approach is applied to calculate the energy-momentum tensor in the Schwarzschild spacetime to describe a tensor distribution in the singular region. From a physical point of view this allows us to identify the matter source for the Schwarzschild geometry. These ideas have been developed later in many works, see, for example, \cite{Pantoja_Rago_2002,Heinzle_Steinbauer_2002} and references there in.

In  \cite{Pantoja_Rago_2002}, the authors have proposed a restrictive kind of regularization inspired by approaches used to the study of the classical gravitational self-energy. As a result, the minimum extension associates to point-like sources in GR described with making the use of the $\delta$-function. It is shown that this approach may be used to regularize non-regular metrics in such a way that the regularized metrics allow us to construct well defined distributional curvature tensors. The curvature and Einstein tensors of the Schwarzschild spacetime are considered as an application. The regularized metric becomes continuous regular metric in the sense of Geroch and Traschen \cite{Geroch_Traschen_1987} with well defined distributional curvature tensor (respectively, with well defined distributional matter energy-momentum tensor) at all the intermediate steps of calculations.

The work \cite{Heinzle_Steinbauer_2002} is devoted to a mathematical analysis of the distributional Schwarzschild geometry specially. The Schwarzschild solution is extended to include the singularity; the energy-momentum becomes a $\delta$-distribution supported at r = 0. Using generalized distributional geometry in the sense of Colombeau’s \cite{Colombeau_1983} construction, the nonlinearities are treated in a mathematically rigorous way. Generalized function techniques are used as a tool to give a unified discussion of various  approaches taken in the literature earlier. It is noted that if a regularization is provided with making the use of the Schwarzschild coordinates then it is either non-smooth or not invertible. Keeping this in mind, it is shown that the Eddington-Finkelstein coordinates are the most preferable ones, they are actually the only ones where Colombeau's construction for the sources successfully works.

In \cite{Fiziev_2004}, utilizing various gauges of the radial coordinate, a description of static spherically symmetric space-times with a point singularity at the center and vacuum outside the singularity is given. Boundary conditions differ significantly from those for the Schwarzschild solution. As a result, new solutions differ from the Schwarzschild solution itself. In GR, there exists a two-parameters family of such new solutions to the Einstein equations which are physically distinguishable but only some of them describe the gravitational field of a single massive point particle with nonzero bare mass $m_0$. Novel normal coordinates and a new physical class of gauges are proposed, achieving a correct description of a point mass source in GR.

In \cite{Goswami_Joshi_Vaz_Witten_2004}, a class of spherically symmetric collapse models in which a naked singularity may develop as the end state of a collapse is constructed. The matter distribution considered has negative radial and tangential pressures, but the weak energy condition is obeyed throughout. The singularity forms at the center of the collapsing cloud and continues to be visible for a finite time. The duration of visibility depends on the nature of energy distribution. Hence the causal structure of the resulting singularity depends on the nature of the mass function chosen for the cloud. A general model, in which the naked timelike singularity is formed, is presented.

In \cite{Cadoni_Magnemi_2005}, two-dimensional solutions of a generic dilaton gravity model coupled with matter, which describe D-dimensional static black holes with point-like sources are derived.

In \cite{Lundgren_Schmekel_York_2007}, a problem of the gravitational binding energy of point-like particles, that diverges even in Newtonian gravity, is considered. In GR, the analog of a point particle is a black hole and the notion of binding energy is suggested to be replaced by quasi-local energy. The quasi-local energy derived by York, and elaborated by Brown and York \cite{Brown_York_1993}, is finite outside the horizon. The authors present a prescription for finding the quasi-local energy inside a horizon, and show that it is finite at the singularity for a variety of types of black hole. The energy is typically concentrated just inside the horizon, but not at the central singularity!?

In \cite{Katanaev_2013}, it is proven that the Schwarzschild solution in the isotropic coordinates satisfies the system of Einstein’s and geodesic equations for a point massive particle. However, the spacetime in such a picture is described by two sheets. As a result, $\delta$-function does not correspond to the true singularity.

All of the above works (and references there in) examine the problem of the point mass in GR and in some gravitational theories as a {\em particular problem}. This can be a specific point-like solution obtained with special boundary conditions, or a represented black hole solution, where singularities are described with making the use of special mathematic techniques, etc. A collapse to a point-like stage is considered under very special equations of state. However, there is no an unique and complex model representing a black hole as point particle including a continues collapse to it that satisfies a set of necessary physically reasonable requirements. Here, exploring the Schwarzschild black hole solution in GR, we present such a description. To realize this goal we restrict ourselves by the following requirements:
\bit

\item[(i)] The true singularity has to be described by the world line $r=0$ with making the use of Dirac's $\del({\bm r})$-function.

\item[(ii)] The Schwarzschild solution has to be presented in the asymptotically flat form with an appropriate fall-off of potentials at spatial infinity consistent with the Newtonian behavior.

\item[(iii)] Such a model has to be consistent with a spherically symmetric collapse process. Therefore, we require that test particles fall from infinity to achieve the true singularity continuously in a spacetime diagram.
\eit
The requirement (i) meets special difficulties discussed around Eqs (\ref{a-4})-(\ref{a-5}) if the standard {\em geometrical} presentation of GR is used. Indeed, then the true singularity presented by the $\del({\bm r})$-function is defined for the timelike coordinate $r$ under the horizon. It is a main reason why in the present paper we apply the field-theoretical formulation of GR (see Sect. \ref{sec:2}), where $r$ is a space-like coordinate of a background spacetime. We have to note that an auxiliary background metric is a necessary part of the approaches developed in \cite{Pantoja_Rago_2002,Heinzle_Steinbauer_2002}, where the powerful and elegant mathematics is explored.

Then, in the framework of the field-theoretical formulation, one can add the following.
\bit

\item[(iv)] We require a so-called ``$\eta$-causality''
(property, when the physical light cone is inside the background light
cone) at all the points of the background spacetime.

\eit
This requirement is necessary to avoid interpretation difficulties under the field-theoretical
presentation of GR. The requirement (iv) means that all of the causally connected
events in the physical spacetime are to be described by the right causal
structure of the background spacetime. Properties of the $\eta$-causality and
gauge transformations conserving it were studied in \cite{Pitts_Schive_2004}. However, this requirement is not necessary, because the background spacetime brings an auxiliary character.

At last, it is not necessary but desirable
\bit
\item[(v)] to require a finite time for a free test particle in the background spacetime to achieve the true singularity.
\eit

The existence of the energy-momentum tensor (not pseudotensor) for
the gravitational field and its matter sources is one of the
advantages of the field-theoretical formulation. This was the main reason why
this formulation was used in \cite{Petrov_Narlikar_1996} to consider the Schwarzschild
solution in the Schwarzschild coordinates (frame) as a gravitational field configuration in a
background Minkowski space. The concept of Minkowski
space was extended from spatial infinity (frame of reference of a
distant observer) up to the horizon $r = r_g$, and even under the horizon including the worldline $r
= 0$ of the true singularity. Then, the energy-momentum tensor was
constructed, the energy distribution and the total energy with
respect to the background were obtained. The configuration satisfies
the Einstein equations at all the points of the Minkowski space,
including $r = 0$. The energy distribution is presented by an
expression proportional to $\del({\bf r})$ and by free gravitational
field outside $r = 0$. Thus, the requirement (i) is satisfied. In spite of advantages, the
interpretation of the point mass in \cite{Petrov_Narlikar_1996} has open questions. At
$r= r_g$ both the gravitational potentials and the energy density
have discontinuities. As a result, the requirement (iii) is lost because in the Schwarzschild frame a description of the ingoing geodesics has discontinuities as well.

Thus, one needs to find a more appropriate frame for Schwarzschild solution and a related field configuration that satisfies all the requirements (i)-(v). Already in \cite{Petrov_2005} we have suggested such a frame, it is the contracting Eddington-Finkelstein (EF) coordinates for that the related field configuration satisfies all the requirements (i)-(v). However, it is only a particular case and a collapse to a point-like object did not considered. In the present paper, we use the EF frame as a basis\footnote{ It is a place to note that in \cite{Pantoja_Rago_2002,Heinzle_Steinbauer_2002} it was clarified that the EF coordinates are the most preferable in their approach as well.} to generalize a description of a point-like model satisfying (i)-(v), and including a collapse stage as well.

\section{Elements of the field-theoretical formulation of GR and preliminaries}
 \m{sec:2}

 \subsection{Gravitational field-theoretical equations}
 \m{3:1}

The field-theoretical formulation was developed in
\cite{GPP_1984,Popova_Petrov_1988,Petrov_1993,Petrov_2008} and is based on the famous paper by
Deser \cite{Deser_1970}, who has generalized the results of previous authors. We briefly repeat the main notions of this approach in \cite{GPP_1984}. Let the Einstein theory is described by the Lagrangian:
 \be {\cal L}= {\cal L}(g^{\mu\nu},\,\phi^A)
=-\frac{1}{2\kappa}{\sqrt{-g}}R(g_{\mu\nu}) + {\cal L}^M(g_{\mu\nu},\,\phi^A),
 \m{Lagrangian}
\ee
where $\phi^A$ is a set of tensor densities (matter fields). Variation of (\ref{Lagrangian}) with respect to $g^{\mu\nu}$ leads the Einstein equations in the usual form (\ref{a-3}). To represent GR in the field-theoretical form one has to decompose the metric, $g_{\mu\nu}$, into the background, $\gamma_{\mu\nu}$, and perturbed (dynamical) parts $h^{\mu\nu}$. A more appropriate way to make it is
 \be
\sqrt{-g}g^{\mu\nu} \equiv {\bm\gamma}^{\mu\nu} + \goh^{\mu\nu} \equiv  \sqrt{-\gamma}\l(\gamma^{\mu\nu} + h^{\mu\nu}\r),
 \m{identification}
 \ee
 where $g \equiv \det {g_{\mu\nu}}$ and $\gamma \equiv \det {\gamma_{\mu\nu}}$, and a concrete coordinate chart $\{x^\alpha\}$ is used. Thus, $h^{\mu\nu}$, is interpreted as a {\em field configuration} propagating in a background spacetime with the metric $\gamma_{\mu\nu}$. Here, it is enough to consider Ricci-flat backgrounds with the dynamical fields $\goh^{\mu\nu}$ and $\phi^A$ dynamics of which are described by the Lagrangian:
 \be
{\cal L}^{dyn}  = {{\cal L}}\l({\bm\gamma} + \goh, \,\phi\r)
 - \goh^{\mu\nu} {{\delta \bar {{{\cal L}}}}\over{\delta {\bm\gamma}^{\mu\nu}
}} - \bar{{{\cal L}}}\,, \m{Ldyn}
\ee
where $\bar{{{\cal L}}} = {\cal L}_{h,\phi = 0}$. Its variation with respect to $\goh^{\mu\nu}$ leads to the gravitational field equations:
 \be
G^L_{\mu\nu}(h^{\alpha\beta}) = {\kappa {t^{tot}_{\mu\nu}}}\, ,
 \m{FieldEqs}
 \ee
which are equivalent to the Einstein equations in the usual form (\ref{a-3}). The left hand side is linear in $h^{\mu\nu}$:
 \be
G^L_{\mu\nu}(h^{\alpha\beta}) \equiv {\half
\l(h^{~~~;\alpha}_{\mu\nu~~;\alpha} +
\gamma_{\mu\nu}h^{\alpha\beta}_{~~~;\alpha\beta} -
h^\alpha_{~\mu;\nu\alpha} - h^\alpha_{~\nu;\mu\alpha}\r)}\,,
 \m{GL}
 \ee
where   ${}_{;\alpha}$ means the covariant derivative with respect to $\gamma_{\mu\nu}$. The total
energy-momentum tensor
 \be
t^{tot}_{\mu\nu} \equiv {t^g_{\mu\nu} + t^m_{\mu\nu}}
 \m{t-tot}
 \ee
is obtained after varying the Lagrangian (\ref{Ldyn}) with respect to $\gamma^{\mu\nu}$:
\bea
{t}^{\rm tot}_{\mu\nu} &\equiv & \frac{2}{\sqrt{-\gamma}}{{\delta{\cal
L}^{dyn}}\over{\delta \gamma^{\mu\nu}}}\,,
  \m{tei}\\
  {t}_{\rm tot;\nu}^{\mu\nu} & = & 0\,.
 \eea
The pure gravitational part of
(\ref{t-tot}) has the form:
 \be
\kappa t^g_{\mu\nu} = -(\Del\Del)_{\mu\nu} + \half \gamma_{\mu\nu}
(\Del\Del)^{~\alpha}_\alpha + Q^\sigma_{~\mu\nu;\sigma}
 \m{t-g}
 \ee
 with the tensors
 \bea
 (\Del\Del)_{\mu\nu}& \equiv &{\Del^\alpha_{~\mu\nu}\Del^\beta_{~\beta\alpha} -
 \Del^\alpha_{~\mu\beta}\Del^\beta_{~\nu\alpha}}\, ,
 \m{KK}\\
Q^{\sigma}_{~\mu\nu}& \equiv & {-
\half\gamma_{\mu\nu}h^{\alpha\beta} \Del^\sigma_{~\alpha\beta}} + \half
{h_{\mu\nu}\Del^{\alpha~\sigma}_{~\alpha}}
 - {h^{\sigma}{}_{(\mu}\Del^{\alpha}_{~\nu)\alpha}}\nonumber\\ & +&
h^{\beta\sigma}\Del^\alpha_{~\beta(\mu}\gamma_{\nu)\alpha} +
h^{\beta}{}_{(\mu}\Del^\sigma_{~\nu)\beta} - h^{\beta}{}_{(\mu}
\gamma_{\nu)\alpha}\Del^\alpha_{~\beta\rho}\gamma^{\rho\sigma}\, ,
 \m{Q}\\
\Del^\alpha_{~\beta\gamma} &\equiv & \Gamma^\alpha_{~\beta\gamma} -
 C^\alpha_{~\beta\gamma}\,
 \m{K}
 \eea
where  $\Gamma^\alpha_{~\beta\gamma}$ and $C^\alpha_{~\beta\gamma}$
are the Christoffel symbols for the dynamic (physical) and
background spacetimes respectively. Note that really the quantity (\ref{K}) is defined by the components $h^{\mu\nu}$.
The matter energy-momentum tensor, $t^m_{\mu\nu}$, in (\ref{t-tot})  is connected by a special way with the usual
matter energy-momentum tensor $T_{\mu\nu}$ of GR in (\ref{a-3}):
 \be
t^m_{\mu\nu} = T_{\mu\nu} - \half
g_{\mu\nu}T_{\alpha\beta}g^{\alpha\beta} - \half
\gamma_{\mu\nu}\gamma^{\alpha\beta}\l(T_{\alpha\beta} - \half
g_{\alpha\beta}T_{\pi\rho}g^{\pi\rho}\r)\, .
 \m{t-T}
 \ee

It can be more economical if we do not calculate concrete sources, $t^{tot}_{\mu\nu}$. Instead of that, assuming that the field equations (\ref{FieldEqs}) hold, we can find components of $t^{tot}_{\mu\nu}$ equalising them to the left hand side in (\ref{FieldEqs}):
 \be
{{t^{tot}_{\mu\nu}}} = \kappa^{-1}G^L_{\mu\nu}(h^{\alpha\beta})\, .
 \m{FieldEqs+}
 \ee

 \subsection{Gauge transformations}
 \m{3:2}

 The important here properties of gauge transformations have to be remarked. {\em The same } solution to the Einstein equations can be written in another coordinate chart, say, $\{x'^\alf\}$. The corresponding decomposition is $\sqrt{-g'}g'^{\mu\nu}(x') \equiv  \sqrt{-\gamma'}\l(\gamma'^{\mu\nu}(x') + h'^{\mu\nu}(x')\r)$. Then, after the shifting in the frame $\{x'^\alf\}$ from points with values of the coordinates $x'^\alf$ to points with values $x^\alf$ and after equalizing $\gamma'^{\mu\nu}(x) = \gamma^{\mu\nu}(x)$, one gets
 \be
\sqrt{-g'}g'^{\mu\nu}(x) \equiv  {\bm\gamma}^{\mu\nu}(x) + \goh'^{\mu\nu}(x) \equiv  \sqrt{-\gamma}\l(\gamma^{\mu\nu}(x) + h'^{\mu\nu}(x)\r)
 \m{identification+}
 \ee
 instead of (\ref{identification}). A connection of (\ref{identification}) and (\ref{identification+}) and its interpretation is as follows. They are related to the same solution to the Einstein equations; for both of these decompositions the same background presented by the metric $\gamma_{\mu\nu}$ is chosen by different ways; thus, one concludes that the fields $h^{\mu\nu}$ and $h'^{\mu\nu}$ describe the same physical reality, only they are connected by gauge transformations, see \cite{GPP_1984,Petrov_2008}. Symbolically, such transformations, including the matter variables are
 \bea
\goh'^{\mu\nu}(x) & = & \goh^{\mu\nu}(x) + \delta\goh^{\mu\nu}(x);
 \m{gauge_h}\\
 \phi'^{A}(x) & = & \phi^{A}(x) + \delta\phi^{A}(x).
 \m{gauge_phi}
 \eea
 Here, it is important to note that the field equations (\ref{FieldEqs}) are gauge-invariant on the background equations and themselves.

Let us illustrate transformations (\ref{gauge_h}) and (\ref{gauge_phi}) more. The aforementioned shift is called usually the Lie displacement. Defining a tangential vector to a congruence of this displacement as $\xi^\mu$, we can rewrite (\ref{gauge_h}) and (\ref{gauge_phi} in the form of expansions:
 \bea
 {\goh}'^{\mu\nu} &=&  \goh^{\mu\nu} +
\sum^{\infty}_{k = 1}{1\over{k!}}~ \hbox{$\pounds$}_\xi^k \l({\bm\gamma}^{\mu\nu} + \goh^{\mu\nu}\r), \nonumber\\ {\phi}'^A &=& \phi^A + \sum^\infty_{k =
1}{1\over{k!}}~\hbox{$\pounds$}_\xi^k \phi^A.
\nonumber
 \eea
These gauge transformations in linear gravity theory on a flat background in Lorenzian coordinates acquire the well known form, see in \cite{Petrov+Co_2017}:
 \be
\goh'^{\mu\nu} = \goh^{\mu\nu} +
\hbox{$\pounds$}_\xi {\bm\eta}^{\mu\nu}, ~~{\rm or}
~~h'^{\mu\nu} = h^{\mu\nu}  - \eta^{\mu\nu}\di_\rho \xi^\rho
+ \di^\mu \xi^\nu + \di^\nu \xi^\mu
 \m{linearGRgauge}
 \ee
 Below we will use the transformations in the exact and close form (\ref{gauge_h}) only.

\subsection{Preliminaries}
 \m{3:3}

 In the present paper, the given above properties of the field-theoretical method are used to describe the Schwarzschild black hole as point particle.

 We consider a general form for the metric of the Schwarzschild solution as
 \be
 ds^2 =-\omega(r_g/r) c^2dt^2 + 2\varkappa(r_g/r)cdtdr + \varrho(r_g/r)dr^2 +  r^2 d\Omega^2\,
 \m{S_G}
 \ee
 A flat (auxiliary) background spacetime is chosen in the form
 \be
 d\bar s^2 =- c^2dt^2 + dr^2 +  r^2 d\Omega^2\,.
 \m{fon}
 \ee
 Using the recipe (\ref{identification}), we represent (\ref{S_G}) as field configuration, $h^{\mu\nu}(x)$, in Minko\-w\-ski space with the metric (\ref{fon}).

 If a one of the properties of $h^{\mu\nu}(x)$ does not satisfy a requirement from the set (i)-(v) in Sect. \ref{Preliminaries}, it can be improved with making the use of the gauge transformations (\ref{gauge_h}). We apply coordinate transformations of the type
 \be
 cdt' =cdt + \psi(r_g/r)dr\,
 \m{t_TO_t}
 \ee
 only. After that the metric transforms to
 \be
 ds^2 =-\omega(r_g/r) c^2dt'^2 + 2\varkappa'(r_g/r)cdt'dr + \varrho'(r_g/r)dr^2 +  r^2 d\Omega^2\,.
 \m{S_G+}
 \ee
Providing the Lie displacement and, thus, exchanging $t'$ in (\ref{S_G+}) with $t$, choosing again the background metric in the form (\ref{fon}), we calculate $h'^{\mu\nu}(x)$ with the use of the decomposition (\ref{identification+}). Of course, the field configurations, $h'^{\mu\nu}(x)$ and $h^{\mu\nu}(x)$, are connected by the gauge transformations (\ref{gauge_h}). Let us shortly discuss the requirements (i)-(v) in Sect. \ref{Preliminaries}.

First, the requirement (i) takes on a place if the field configuration, $h^{\mu\nu}(x)$, used in (\ref{GL}) defines the energy-momentum distribution in (\ref{FieldEqs+}) with $\delta$-function in the region $r=0$ of the background spacetime (\ref{fon}). To calculate it carefully one has to use the technique of the generalized functions \cite{Gelfand_Shilov_1958}. Here, it is important to define the expression
$\nabla^2(1/r^{k+1})$ with integer $k \ge 0$, for which we obtain the final
expression,
 \be
 \nabla^2\frac{1}{r^{k+1}}= (k+1)\l[ \frac{k}{r^{k+3}}
-\frac{4\pi}{r^{k}}  \delta({\bf r})\r]\, .
 \label{Delta-rk++}
 \ee
It is easy to see that integration over a round ball of the right hand side
of (\ref{Delta-rk++}) gives two divergent integrals at $r \goto 0$
that compensate one another. Then, a convergent part of this volume
integral is equal to a value of a surface integral that follows
after integration of the left hand side of (\ref{Delta-rk++}), which is a
divergence $\nabla^2 = \di_i\di^i$.

Second, for a spherically symmetric static system among all the integral conserved quantities only the total energy does not vanish. From a one hand, it can be calculated making the use of the volume integration,
 \be
 E = \lim_{r \goto \infty} \int_V t^{tot}_{00}r^2\sin\theta dr
 d \theta d \phi\, .
 \m{int-V}
 \ee
However, due to the field equations (\ref{FieldEqs+}) and the explicit expression (\ref{GL}) the volume integration can be replaced with the surface integration over the 2-sphere with $r = r_0$:
 \be
 E = \lim_{r_0 \goto \infty}\frac{1}{2\k} \oint_{\di V}
 \l(h_{00}{}^{;1} + \gamma_{00} h^{1\alpha}{}_{;\alpha} - 2h^{1}{}_{0;0} \r)
 r^2\sin\theta d \theta d \phi\, .
 \m{int-diV}
 \ee
 We remark that (\ref{int-diV}) gives the acceptable result $E = Mc^2$ if the asymptotic of the field configuration, $h^{\mu\nu}$, corresponds to the Newtonian one $1/r$. It is the sense of the requirement (ii).

 However, it is well known that the definition of the conserved quantities for isolated systems significantly depends on an asymptotic behavior metric coefficients. It turns out that the standard behavior at spatial infinity $\sim 1/r$ can be significantly weakened. For the weakest fall-off see \cite{Bizon_Malec_1986,OMurchadha_1986,Kennefick_OM_1995,Petrov_2008} and references therein. Besides, the same restrictions follow if one tries to present self-consistent asymptotic Poincar\'e algebra \cite{Soloviev_1985} for asymptotically flat spacetimes.

In the case of the field-theoretical presentation, the weakest fall-off has been formulated for the asymptotic behaviour of the field configuration $h^{\mu\nu}$, see \cite{Petrov_1995,Petrov_1997}, as well. To derive them for the Schwarzschild solution one has to transfer to asymptotic Cartesian coordinates, then the background metric (\ref{fon}) acquires the Minkowski form, whereas in these coordinates the asymptotic behaviour for the field components has to be
 \be
 h^{\mu\nu} \sim 1/r^{\alf}~~ {\rm with} ~~\alf > 1/2.
 \m{h_weakest}
 \ee
 For such a behaviour one has again $E = Mc^2$, and the requirement (ii) can be weakened by (\ref{h_weakest}) as well. Note that because (\ref{int-V}) and (\ref{int-diV}) are equivalent, for defining an acceptable mass of the system it is enough the acceptable asymptotic behaviour, it is not necessary to clarify the intrinsic structure of an isolated object.

 Third, considering test particles falling into the true singularity, we turn to solutions for geodesics for the metric (\ref{S_G}) and study their trajectories on the $t\times r$ diagram. If the trajectory does not satisfy the requirement (iii), then we provide an appropriate transformation of the type (\ref{t_TO_t}) to improve the trajectory. Examining the requirement (iii), it is not necessary to use the field-theoretical framework because a coordinate diagram $t\times r$ for (\ref{S_G}) and the frame of the background spacetime (\ref{fon}) coincide.

 Forth, following the requirement (iv) we have to compare the light cones for the physical spacetime (\ref{S_G}) and for the background spacetime (\ref{fon}). We derive related expressions solving the equations $ds=0$ and $d\bar s = 0$ for the radial light rays.

 Fifth, the requirement (v) can be examined by studying the limits at $r\goto 0$ of both geodesics and physical light cones.

 From the start we consider the Schwarzschild solution (\ref{S_G}) in the Schwarzs\-child coordinates (\ref{a-2}) and find out that not all of the requirement are satisfied. Then the Eddington-Finkelstein (EF) coordinates and the related gauge fixing turn out the most appropriate ones to satisfy all the requirements (i)-(v). After that, basing on the EF coordinates, we generalize the gauge fixing being restricted maximally only by the requirements (i)-(v).

 Analysis of a spherically symmetric collapse is divided into examining the extrinsic region (that exactly coincides with the vacuum case) and examining the the intrinsic region presented by the dust matter. Both regions are described in the unique gauge fixing.

\section{A black hole solution as a field configuration with the $\delta$-function}
 \m{FC-delta}

\subsection{The Schwarzschild gauge fixing}
 \m{S}

  Let us turn firstly to the solution (\ref{a-2}). In \cite{Petrov_Narlikar_1996}, with the use of the field-theoretical approach it has been shown that, analogously to the Newtonian prescription (\ref{a-1}), the Schwarzschild solution in the form (\ref{a-2}) can be described as a point mass in GR  in a non-contradictory manner. In such a picture, a flat (auxiliary) background spacetime with the metric (\ref{fon}) is identified with asymptotically flat spacetime of the solution (\ref{a-2}). Thus, we exploit the model, when the spacetime of the solution (\ref{a-2}) and the background Minkowski space are in a one-to-one correspondence. The related field configuration, $h^{\mu\nu}_s$, defined with the use (\ref{identification}) is
 \be
 h_s^{00}  = - \frac{{r_g}/{r}}{1- {r_g}/{r}}\,
 ,\qquad  h_s^{11}  =  -\frac{r_g}{r}\,.
  \m{h-configur-S}
 \ee

 {\em The requirement (i):} To calculate the components of the energy-momentum tensor (\ref{t-tot}), $t^{tot}_{\mu\nu}$, for such a configuration we use the convention (\ref{FieldEqs+}) at all the points of the Minkowski space. Applying this technique (\ref{Delta-rk++}) for calculating the right hand side in (\ref{FieldEqs+}), one obtains for these components at $r = 0$ that they are proportional to $\del({\bm r})$, see \cite{Petrov_Narlikar_1996}. Thus, the requirement (i) is satisfied.

  {\em The requirement (ii):} The field configuration (\ref{h-configur-S}) has just the Newtonian asymptotic and its substitution into (\ref{int-diV}) gives the acceptable result $E = Mc^2$.  Thus, the requirement (ii) is satisfied.

{\em The requirement (iii):} Reflecting the coordinate singularity at $r = r_g$ in the Schwarzschild coordinates in (\ref{a-2}), the geodesics have a break at the horizon on the $t\times r$ diagram, see textbook \cite{Landau_Lifshitz_1975}. Let us present the well known textbook formulae, which are the basis for our below study.

  To simplify the presentation we consider test particles, falling radially into a black hole. Besides, we restrict ourselves to the ``parabolic orbit'' case, when a particle begins its motion from the rest at the infinity $r = \infty$. To derive such trajectories of test particles one has to resolve the geodesic equations \cite{Landau_Lifshitz_1975} for the solution  (\ref{a-2}). As a result we obtain the 4-velocity, $u_s^\alf$:
 \be
  u_s^0 =
\frac{1}{1- r_g/r}\, ,\qquad u_s^1 =
-\l(\frac{r_g}{r}\r)^{1/2} , \qquad u^2=u^3 = 0.
\m{c-d71}
\ee
After integration of $cdt =(u_s^0/u_s^1)dr$ one obtains the equation of the trajectory on the spacetime, $t\times r$, diagram:
\bea ct & = & -
r_g\l[{{ \frac{2}{3}}} \l({\frac{r_g}{r}}\r)^{-3/2}+
2\l(\frac{r_g}{r}\r)^{-1/2} \r.
\nonumber \\
& {} & + \l.\ln \l|{\frac{r}{r_g}- 1}\r| -
2\ln \l|{\l({\frac{r_g}{r}}\r)^{-1/2} +1}\r|\,\r]
+ {\rm const}\,  .
\m{c-d61}
\eea
The existence of the term $-r_g \ln |r/r_g - 1 |$ leads to the situation, when a particle falls to the event horizon $r = r_g$ infinitely long in the coordinate time $t$, the same in the coordinates of the background (\ref{fon}). This means that the requirement (iii) is not satisfied.

{\em The requirement (iv):} Let us derive the expression defining the light cone on the spacetime, $t\times r$, diagram for the solution (\ref{a-2}):
\be
\l.\frac{cdt}{dr}\r|_{1,2} = \pm \frac{1}{1 - {r_g}/{r}}.
\m{S-cone}
\ee
One can easily recognize that the expression (\ref{S-cone}) signals that the requirement (iv) is not hold inside horizon. Besides, analyzing  (\ref{S-cone}) one can conclude that a geodesic at $r = r_g$ has a break because the light cone is degenerated at the event horizon.

{\em The requirement (v):} Inside the event horizon the expression (\ref{S-cone}) tells us that the point particle moves in unusual way in time $t$ of the background Minkowski space. Therefore, the requirement (v) is not hold as well.

\subsection{The Eddington-Finkelstein gauge fixing}
 \m{EF_fixing}

The problem of the break at $r= r_g$ has to be countered with the use of an appropriate choice of coordinates that can help us to reformulate the field configuration. At least, one could use the coordinates without singularities at the horizon, like Novikov's, Kruskal-Szekeres's, {\it etc.}, coordinates \cite{Landau_Lifshitz_1975,Misner_Thorn_Wheeler_1973}, which can resolve the problem locally at neighborhood of $r= r_g$. However,  many forms of the Schwarzschild solution in these coordinates do not satisfy the requirement (ii), or they cover the whole Schwarzschild geometry that contains causally not connected domains that is not desirable. Among them, the use of the Eddington-Finkenstein (EF) coordinates in stationary form \cite{Eddington_1924,Finkelstein_1958,Misner_Thorn_Wheeler_1973} resolves these problems, see \cite{Petrov_2005}, and all of the requirements (i)-(v) are satisfied. Here, it is quite instructive to repeat in a more detail these results and develop them below.

Transformation to the EF coordinates from the Schwarzschild coordinates in (\ref{a-2}) is as follows,
\be
cdt_e = cdt_s + \frac{r_g/r}{1 - {r_g}/{r}}dr = cdt_s + \l(\frac{1}{1 - {r_g}/{r}} - 1 \r)dr.
\m{EF-transformations1}
\ee
Then, the contracting EF metric for
the Schwarzschild geometry is
 \be
 ds^2 = - \l(1 - \frac{r_g}{r}\r)c^2dt_e^2 + 2\,\frac{r_g}{r}cdt_e dr
+ \l(1 + \frac{r_g}{r}\r)dr^2 +
 r^2 d\Omega^2 \,.
 \m{EF}
 \ee
 To make a gauge transformation from $h^{\mu\nu}$ to $h'^{\mu\nu}$ in (\ref{gauge_h}) one has to provide a shift $t' \goto t$. In the present case one  has to provide a shift $t_e \goto t$. Then, choosing the flat background again in the form (\ref{fon}), and making the use of the decomposition (\ref{identification+}) the
field configuration corresponding to (\ref{EF}) is obtained,
 \be
 h_e^{00}  =  - \frac{r_g}{r}\, ,\qquad h_e^{01}  =   \frac{r_g}{r}\, ,\qquad
 h_e^{11}  =  -\frac{r_g}{r}\, .
  \m{h-configur-EF}
 \ee

{\em The requirement (i).} The field configuration (\ref{h-configur-EF}) has to satisfy the Einstein equations (\ref{FieldEqs}) in the whole Minkowski space, including the world line $r=0$. We apply the technique (\ref{Delta-rk++}) to calculate the components of $t^{tot}_{\mu\nu}$ defined in (\ref{FieldEqs+}). As a result, we obtain non-vanishing components of the total energy-momentum tensor:
\bea
 t^{tot}_{00} & = & Mc^2 \delta({\bm r})\, ,\nonumber\\
t^{tot}_{11} & = & - Mc^2\delta({\bm r})\, ,\nonumber\\
t^{tot}_{AB} & = & - \half \gamma_{AB}\, Mc^2\delta({\bm r}); \qquad A,\ldots = 2,3\,.
 \m{t-tot-EF}
 \eea
One can see that the energy-momentum is concentrated {\it
only} at $r=0$, and it is expressed with making the use of the $\delta({\bm r})$-function. Thus the requirement (i) is satisfied.

\noindent {\em The requirement (ii).} The integration (\ref{int-V}) of $t^{tot}_{00}$ in (\ref{t-tot-EF}) determined by the $\delta({\bf r})$-function only, gives $E = Mc^2$. Of course, (\ref{int-diV}) leads to the same acceptable result because the asymptotic of (\ref{h-configur-EF}) corresponds to the description of the asymptotically flat spacetime as well. Thus, the requirement (ii) is satisfied. It is interesting to note that unlike  of many different situations, here  $Mc^2$ follows with an arbitrary radius of the 2-sphere $r_0$ (it is not necessary $r_0 \goto \infty$), like for the electric charge in electrodynamics and for the point mass in Newtonian gravity.

{\em The requirement (iii).} The transformation (\ref{EF-transformations1}) permits us to recalculate
the components of 4-velocity for test particles (\ref{c-d71}) in the EF coordinates:
 \be
  u_e^0 = 1+
\frac{r_g/r}{1+ (r_g/r)^{1/2}}\, ,\qquad u_e^1 =
-\l(\frac{r_g}{r}\r)^{1/2} , \qquad u^2=u^3 = 0.
\m{c-d71_a}
\ee
After integration of $cdt =(u_e^0/u_e^1)dr$ one obtains the equation of the radial parabolic orbits on the spacetime, $t\times r$, diagram:
\be
ct = -
2r_g\l[{{ \frac{1}{3}}} \l({\frac{r_g}{r}}\r)^{-3/2}+
\l(\frac{r_g}{r}\r)^{-1/2}  -
\ln \l|{\l({\frac{r_g}{r}}\r)^{-1/2} +1}\r|\,\r]
+ {\rm const}\,  .
\m{c-d61+}
\ee
Such trajectories are continuous up to the true singularity $r=0$. Thus the requirement (iii) is satisfied as well.

{\em The requirement (iv).} Analyzing the light cone for the solution (\ref{EF}),
\be
\l.\frac{cdt}{dr}\r|_{1} = - 1,\qquad \l.\frac{cdt}{dr}\r|_{2} =  \frac{1+ {r_g}/{r}}{1 - {r_g}/{r}},
\m{EF-cone}
\ee
we see also that it is inside the light cone of the Minkowski space everywhere. Thus, the requirement (iv) is satisfied. Besides, the cone (\ref{EF-cone}) is not degenerated at the event horizon.

\noindent {\it The requirement (v).} The behaviour of the cone (\ref{EF-cone}) at the limit $r\goto 0$ tells us that particles achieve the true singularity in a finite time $t$ of the Minkowski space.

\subsection{Generic gauge fixing}
 \m{G_generic}

 The goal of the present subsection is to generalize the results of previous subsection. Again to construct a generalized configuration, developing  (\ref{h-configur-EF}), we provide coordinate transformations of the type (\ref{t_TO_t}). If one begins from the Schwarzschild solution in the Schwarzschild coordinates (\ref{a-2}) one has to again escape the break of geodesics at $r = r_g$ including into the transformation for a new $dt_f$  the term $\l(1 - r_g/r \r)^{-1}dr$, like in (\ref{EF-transformations1}). Therefore we begin the generalization from the EF frame developed in the previous subsection and provide the transformation:
\be
cdt_f = cdt_e + f(r_g/r)dr ,
\m{EF_to_generic}
\ee
with the enough smooth function $f(r_g/r)$ at $0 < r \le \infty$ to conserve the continuity of geodesics on diagrams $t_f\times r$, maybe except of the true singilarity. Then, the EF metric (\ref{EF}) for
the Schwarzschild geometry transforms to the generic form:
 \bea
 ds^2 = &-& \l(1 - \frac{r_g}{r}\r)c^2dt_f^2 + 2\l[\frac{r_g}{r} + \l(1 - \frac{r_g}{r}\r)f\r] cdt_f dr\nonumber\\
&+& \l[\l(1 + \frac{r_g}{r}\r) - 2\frac{r_g}{r}f - \l(1 - \frac{r_g}{r}\r)f^2\r]  dr^2 +
 r^2 d\Omega^2 \,.
 \m{generic}
 \eea
Now, one has to provide a shift $t_f \goto t$.  Then, choosing the flat background again in the form (\ref{fon}), and making the use of the decomposition (\ref{identification+}) the
field configuration corresponding to (\ref{generic}) becomes
 \bea
 h_f^{00}  &=&  - \frac{r_g}{r} + 2\frac{r_g}{r}f + \l(1 - \frac{r_g}{r}\r)f^2\,
 ,\nonumber\\
 h_f^{01}  &=&   \frac{r_g}{r} + \l(1 - \frac{r_g}{r}\r)f\, ,
 \nonumber\\
 h_f^{11}  &=&  - \frac{r_g}{r}\, .
  \m{h-generic}
 \eea

 {\em The requirement (i).} As before, the field (\ref{h-generic}) has to satisfy the Einstein equations (\ref{FieldEqs}) in the whole Minkowski space. Because the field configuration (\ref{h-generic}) contains an arbitrary function $f=f({r_g}/{r})$ the technique has to be generalized. Formally, one can derive
 \be
 \nabla^2 f\l(\frac{r_g}{r}\r) = \l(f''\frac{r^2_g}{r^4} - 4\pi r_g f'\delta(\bm r)\r) ,
 \m{Delta-f}
 \ee
 where $f' = \di_x f(x)$. At least, the definition (\ref{Delta-f}) corresponds to the formula (\ref{Delta-rk++}) with $f = (r_g/r)^n$ with integer $n\geq 0$. Application of the theory of generalized functions requires a quite careful consideration, therefore formula (\ref{Delta-f}) plays a role of the restriction for a choice of $f\l({r_g}/{r}\r)$. Then, calculating the components of $t^{tot}_{\mu\nu}$ with making use of (\ref{FieldEqs+}), we obtain non-vanishing components of the total energy-momentum tensor:
\bea
 t^{tot}_{00} & = & mc^2 \delta({\bm r}) -4\pi r_g \delta({\bm r}) \l[2\l(f + \frac{r_g}{r}f'\r)+ 2 ff' - f^2 - 2\frac{r_g}{r}ff' \r] \nonumber\\ && + \l[4f'^2 + \l(f'' - f'^2\r)\frac{r_g}{r} - 4ff' + ff''\l( 1- \frac{r_g}{r}\r)
  \r]\frac{r^2_g}{r^4}\, ,\m{t-tot-00}\\
t^{tot}_{11} & = & - mc^2\delta({\bm r})\, ,\m{t-tot-11}\\
t^{tot}_{AB} & = & - \half \gamma_{AB}\, mc^2\delta({\bm r}); \qquad A,\ldots = 2,3\,.
 \m{t-tot-AB}
 \eea
One can see that the energy-momentum is concentrated at $r=0$, and it is expressed with making the use of the $\delta({\bm r})$-function. Thus the requirement (i) is satisfied.

{\it The requirement (ii)}. The requirement for the field configuration $h_f^{\mu\nu}$ in (\ref{h-generic}) to have the asymptotic behaviour of the type (\ref{h_weakest}) provides the restriction for the asymptotic behaviour of $f$, namely,
\be
\l.f(r_g/r)\r|_{r\goto \infty} \sim (r_g/r)^{\alf}; \qquad \alf > 1/2.
\m{first-f}
\ee
Indeed, for $f(r_g/r) \sim (r_g/r)^{\alf}$ in the case $\alf = 1/2$ the integrand in (\ref{int-diV}) leads to finite $E \neq Mc^2$;  in the case $\alf < 1/2$ the integrand in (\ref{int-diV}) leads to infinite $E$. Thus, the requirement (ii) gives another restriction (\ref{first-f}) for the function $f$.

{\em The requirement (iii).} The transformation (\ref{EF_to_generic}) permits us to recalculate
the components of 4-velocity for test particles (\ref{c-d71_a}) in the generic coordinates:
 \be
  u_f^0 = 1+
\frac{r_g/r}{1+ (r_g/r)^{1/2}} - f\cdot\l(\frac{r_g}{r}\r)^{1/2} ,~~ u_f^1 =
-\l(\frac{r_g}{r}\r)^{1/2} , ~~ u^2=u^3 = 0.
\m{c-d71_b}
\ee
This, gives
 \be
\frac{cdt}{dr} = \frac{u_f^0}{u_f^1} = - \l({r}/{r_g}\r)^{1/2} -
\frac{(r_g/r)^{1/2}}{1+ (r_g/r)^{1/2}} + f .
\m{du0du1}
\ee
The requirement for geodesics to be continuous after such transformations gives the evident restriction for $f$:
 \be
 \l|f\r| < N
 \m{f<N}
 \ee
with finite arbitrary large positive $N$, at least, for $r>0$. Besides, to have appropriate form of the ingoing geodesics one has to have monotonic smooth function $f$, when ${cdt}/{dr} < 0$. Then, the concrete expression (\ref{du0du1}) gives
 \be
f < 1 + \frac{(r/r_g)^{1/2}}{1+ (r_g/r)^{1/2}}.
\m{du0du1<}
\ee
After integration of (\ref{du0du1}) one obtains the equation of the radial parabolic orbits on the spacetime, $t\times r$, diagram correcting (\ref{c-d61+}):
\bea
ct &=& -
2r_g\l[{{ \frac{1}{3}}} \l({\frac{r_g}{r}}\r)^{-3/2}+
\l(\frac{r_g}{r}\r)^{-1/2}  -
\ln \l|{\l({\frac{r_g}{r}}\r)^{-1/2} +1}\r|\,\r]\nonumber\\
&& + \int^r f\l({\frac{r_g}{r^*}}\r) dr^* +{\rm const}\,  .
\m{c-d61_a}
\eea
Summarizing we conclude that the requirement (iii) is satisfied by the restrictions (\ref{f<N}) and (\ref{du0du1<}) for monotonic smooth $f$ at $0< r \leq \infty$.

{\em The requirement (iv).} Deriving the light cone expressions from $ds^2 =0$ for the quite complicated form of the metric (\ref{generic}), one surprisingly obtains the simple formulae. Thus for the ingoing light ray one has
\be
\l.\frac{cdt}{dr}\r|_{1} = f - 1,
\m{generic-cone1}
\ee
whereas for the outgoing light ray it is
\be
 \l.\frac{cdt}{dr}\r|_{2} = f + \frac{1+ {r_g}/{r}}{1 - {r_g}/{r}}.
\m{generic-cone2}
\ee
The requirement (iv) for (\ref{generic-cone1}) and (\ref{generic-cone2}) can be realized as
\bea
&&f - 1 \leq  -1,\m{generic-cone_x1}\\
&&f + \frac{1+ {r_g}/{r}}{1 - {r_g}/{r}} \geq  1.
\m{generic-cone_x2}
\eea
The restriction (\ref{generic-cone_x1}) gives $f\leq 0$  everywhere $0\leq r \leq \infty$. Then, if we support the requirement (iv) it is not necessary to consider (\ref{du0du1<}). The restriction (\ref{generic-cone_x2}) has to be analyzed in more detail. Considering asymptotic behaviour at $r\goto\infty$ in (\ref{generic-cone_x2}) we are restricted by
\be
\l|f\r|_{r\goto\infty} <~\sim \frac{2r_g}{r}.
\m{generic-cone_a}
\ee
It is stronger than the restriction (\ref{first-f}), therefore if we support the requirement (iv) it is not necessary to take into account (\ref{first-f}). From (\ref{generic-cone_x2}) for the domain $r_g < r < \infty$ one has
\be
\l|f\r| \leq  \frac{2{r_g}/{r}}{1 - {r_g}/{r}}.
\m{generic-cone_b}
\ee
It is the additional restriction to (\ref{f<N}). At last, for the case $r = r_g$ with the restricted $f$, see (\ref{f<N}), the expression (\ref{generic-cone2}) describing  the event horizon becomes $+\infty$ how it has to be. Thus, for a monotonic, restricted and negative $f$ the expression (\ref{generic-cone2}) for the outgoing light ray is positive for $r_g\leq r \leq \infty$.

The case $r < r_g$ requires a special attention. The expression (\ref{generic-cone2}) becomes negative automatically satisfying the requirement (iv) with the natural relation between ingoing and outgoing light rays:
\be
f - 1 \geq f + \frac{1+ {r_g}/{r}}{1 - {r_g}/{r}}.
\m{generic-cone_c}
\ee
The equality in (\ref{generic-cone_c}) takes on a place at the true singularity only as well as only at the true singularity the light cone becomes degenerated. Again, this fact signals on the continuity of the geodesic up to the true singilarity.

It is simply to see that to satisfy the {\em requirement (v)} it is necessary to add the restriction (\ref{f<N}) by
 \be
  \l.\l|f\r|\r|_{r \goto 0} < N
 \m{f<N_0}.
 \ee

Finalizing the section, we repeat that the requirements (iv) and (v) are not so necessary. In fact, one could be restricted by the requirements (i)-(iii) to describe the point-like state of the isolated system in GR by an appropriate way.

\section{Regular gravitational collapse to a point mass}
\m{collapse}

Presenting the Schwarzschild solution as a point particle with making the use of the Dirac $\delta$-function (as it has been demonstrated above), one triggers the question: how can one describe forming the final point mass; what is the way to describe such a collapse? Firstly the gravitational collapse has being studied by Oppenheimer and Snyder \cite{Oppenheimer_Snyder_1939}. More detail on a development of this topic one can find in textbook \cite{Misner_Thorn_Wheeler_1973}. The intrinsic solution presents Friedmann-Robertson-Walker solution with dust in synchronous comoving coordinates; the exterior is presented by the Schwarzschild solution. The main problem is to impose a junction condition to connect smoothly the intrinsic region with the extrinsic region described in different coordinates. However, it looks more natural to describe both of the regions in unique coordinates. Recently such a task has been resolved with the use of a generalization of the well known Painlev\'e-Gullstrand (PG) coordinates in the paper \cite{Kanai_Siino_Hosoya_2011}, elements of which we give below.

\subsection{A continuous gravitational collapse in the Painlev\'e-Gullstrand coordinates}
\m{collapse_in_PG}

The PG coordinates have been discovered independently by Painlev\'e \cite{Peinleve_1921} and Gullstrand \cite{Gullstrand_1922} to represent the Schwarzschild solution. More details on the PG coordinates one can find in \cite{Hamilton_Lisle_2008} and references there in. The original form of the Schwarzschild vacuum solution in the PG coordinates is
 \be
 ds^2 = -c^2dt_p^2 + \l(  dr  + \l({\frac{r_g}{r}}\r)^{1/2} cdt_p \r)^2 + r^2d\Omega^2\,.
\m{PG}
 \ee
 Its main property is that each of sections defined as $ct_p = \const$ presents a flat Euclidean space. Last time the interest to these coordinates arises; many authors, basing on this property,  generalize the PG coordinates for more complicated black holes than the Schwarzschild one, see, for example, \cite{Kanai_Siino_Hosoya_2011,Lin_Soo_2009,Jaén_Molina_2016} and reference there in.

 To obtain the PG coordinates one has to provide the transformations from the Schwarzschild coordinates in (\ref{a-2}) by the way:
\be
cdt_p = cdt_s + \frac{(r_g/r)^{1/2}}{1 - {r_g}/{r}}dr = cdt_s + \l(\frac{1}{1 - {r_g}/{r}} - \frac{1}{1 + (r_g/r)^{1/2}} \r)dr. \m{to_PG}
\ee
Comparing these transformations with the transformations (\ref{EF-transformations1}) one finds the same term $(1 - {r_g}/{r})^{-1}$ that kills the break for the geodesic trajectories on the spacetime diagram. This fact can be stated explicitly after analyzing the components of 4-velocity for test particles. The transformation (\ref{to_PG}) just permits us to recalculate the components of 4-velocity for test particles (\ref{c-d71}) falling radially from infinity in the PG coordinates:
 \be
  u_p^0 = 1\, ,\qquad u_p^1 =
-\l(\frac{r_g}{r}\r)^{1/2} , \qquad u^2=u^3 = 0.
\m{u_PG}
\ee

The authors of the paper \cite{Kanai_Siino_Hosoya_2011} have generalized the vacuum PG solution (\ref{PG}) to the dust case. First, they have assumed that the metric element has the form:
 \be
 ds^2 = -c^2dt_p^2 + \l(  dr  + \sqrt{\frac{2m}{r}\frac{G}{c^2}} cdt_p \r)^2 + r^2d\Omega^2\,,
\m{PG_in}
 \ee
where $m=m(t_p,r)$. Second, the Einstein equations permit to express the matter energy-momentum $T^{(p)}_\nu{}^\mu$ at the right hand side through the function $m(t_p,r)$ unknown from the start:
\bea
\frac{8\pi}{c^2}T^{(p)}_0{}^0 &=& - \frac{2m'}{r^2},\nonumber\\
\frac{8\pi}{c^2}T^{(p)}_0{}^1 &=& \frac{2\dot m}{r^2},\nonumber\\
\frac{8\pi}{c^2}T^{(p)}_1{}^1 &=& - \frac{2m'}{r^2} + \frac{2\dot m}{r^2}\l(\frac{2m}{r}\frac{G}{c^2}\r)^{-1/2}  ,\nonumber\\
\frac{8\pi}{c^2}T^{(p)}_2{}^2 &=& \frac{8\pi}{c^2}T^{(p)}_3{}^3  = - \frac{m''}{r}  + \l(\frac{\dot m}{2r^2} + \frac{\dot m'}{r} \r)\l(\frac{2m}{r}\frac{G}{c^2}\r)^{-1/2} \nonumber\\
&-& \frac{\dot m m'}{2mr}\l(\frac{2m}{r}\frac{G}{c^2}\r)^{-1/2},
\m{EEqs}
\eea
where `prime' means $d/dr$ and `dot' means $d/cdt_p$.
 Third, as usual, in the dust case the matter energy-momentum has the form:
 \be
 T_p^{\mu\nu} = \rho c^2 u_p^\mu u_p^\nu\,,
\m{T_PG}
 \ee
 where for the 4-velocity of matter particles moving radially, it is assumed that : $u_p^\mu = (1, v(t_p,r),0,0)$. Fourth, the requirement of the consistency of the Einstein equations permits to find $v(t_p,r)$. Thus
 \be
  u_p^0 = 1\, ,\qquad u_p^1 =
-\l(\frac{2m}{r}\frac{G}{c^2}\r)^{1/2} , \qquad u^2=u^3 = 0.
\m{u_PG_in}
\ee
Fifth, the integration of the 00-component of the Einstein equations yields the the function
 \be
 m(t_p,r) = 4\pi \int^r_0 \rho(t_p,r) r^2 dr\,,
\m{m_PG}
 \ee
 where it is assumed that $\rho(t,r) =\phi(r)\psi(t)$ and the conditions $\l. m\r|_{r=0} =0$ and $\l. \rho\r|_{t=0} = \infty$ are imposed. Then, the 10-component of the Einstein equations gives:
 \be
 \rho = \frac{1}{6\pi}\frac{c^2}{G}\frac{1}{(ct)^2} \,
\m{rho_PG}
 \ee
 for $-\infty < t \leq 0$. Thus, a combination of (\ref{m_PG}) with (\ref{rho_PG}) gives
 \be
 \sqrt{\frac{2m}{r}\frac{G}{c^2}} = \frac{2}{3}\frac{r}{|ct_p|}\,.
\m{m_PG_ct}
 \ee
 Substitute it into (\ref{PG_in}), one obtains
 \be
 ds^2 = -\l(1- \frac{4}{9}\frac{r^2}{(ct_p)^2} \r)c^2dt_p^2 - \frac{4}{3}\frac{r}{ct_p} dr cdt_p + dr^2 + r^2d\Omega^2\,.
\m{PG_in_tp}
 \ee
 A single non-zero component of the matter energy-momentum tensor and its trace are
\be
T^p_{00} = \frac{c^4}{6\pi G}\frac{1}{(ct_p)^2} ,\qquad T^p = - \frac{c^4}{6\pi G}\frac{1}{(ct_p)^2}.
\m{T_00}
\ee
We remark that (\ref{rho_PG}) as well as (\ref{T_00}) describe a homogeneous distribution of the dust in the intrinsic PG coordinates.

By the above, the intrinsic solution is presented only. However, it is more interesting to describe a collapse of a star with the radius $r = R(t_p)$. Thus, we assume that the solution (\ref{PG_in})-(\ref{T_00}) describes the intrinsic dust region with $r<R(t_p)$, whereas the solution (\ref{PG}) describes the extrinsic vacuum region with $r>R(t_p)$. The total mass $M$ of the star is a constant and is calculated as
  \be
 M = \l. m(t_p,r)\r|_{r=R(t_p)} = \frac{4\pi}{3}R^3 \rho \,.
\m{M_PG}
 \ee
 One has to remark that in such a model the surface of the star is at rest at infinity\footnote{Collapse from a finite radius has been suggested in \cite{Kanai_Siino_Hosoya_2011} as well. However, we do not consider it here because principally it is the same, but formulae are significantly more complicated.} and its radius $R(t_p)$ monotonically decreases to zero as $t_p\goto 0$. Thus, the dust intrinsic region is contracted monotonically to the true singularity.

 It is very important to recall that in \cite{Kanai_Siino_Hosoya_2011} the authors have shown that the intrinsic and extrinsic regions defined in aforementioned way are smoothly matched each other.

 \subsection{A continuous gravitational collapse in the field-theoretical treating}
\m{collapse_in_F-T}

  To describe the continuous gravitational collapse in the framework of the field-theoretical formulation, it seems,  we could directly apply the recommendations of Sect. \ref{sec:2} to the model of previous subsection. However, the extrinsic metric (\ref{PG}) does not satisfy the requirement (ii). Indeed, treating the transformation (\ref{EF_to_generic}) as a transferring from the EF coordinates to the PG coordinates we find that
\be
cdt_p = cdt_e + \frac{({r_g/r})^{1/2}}{1 + ({r_g/r})^{1/2}}dr,
\m{EF_to_PG}
\ee
where, as usual, $r_g = 2MG/c^2$. In the notations of (\ref{EF_to_generic}) it is %
\be
f_p(r_g/r) =  \frac{({r_g/r})^{1/2}}{1 + ({r_g/r})^{1/2}}
\m{EF_to_PG_f}
\ee
that means that the requirement (\ref{first-f}) does not hold.

To satisfy the requirement (ii) for the total model presented by the extrinsic metric (\ref{PG}) and the intrinsic metric (\ref{PG_in_tp}) we can apply the transformation (\ref{EF_to_PG}) to transfer from the PG time $t_p$ to the EF time $t_e$.
Then, the extrinsic metric (\ref{PG}) is transformed to the EF metric (\ref{EF}). Following the field-theoretical prescription, we make shift $t_e \goto t$, obtain the field configuration (\ref{h-configur-EF}) and $t^{tot}_{\mu\nu} =0$  at $r>R(t)$, see (\ref{t-tot-EF}). The next step is to be the field-theoretical reformulation for the intrinsic solution at $r<R(t)$. Because the function $f_p(r_g/r)$ in (\ref{EF_to_PG_f}) is differentiable and monotonic at $0< r \leq \infty$ a smooth matching between extrinsic and intrinsic solutions is preserved. Besides, because the surface $R(t_p)$ goes to zero monotonically  at $t_p \goto 0$, choosing zero constant after integration of (\ref{EF_to_PG}), one easily finds that $R(t_e) \goto 0$, when $t_e \goto t \goto 0$ with the final state (\ref{t-tot-EF}).

Thus, we see that in the case of the EF frame for the intrinsic region the requirements (i)-(iii) and (v) are hold. The requirement (iv) that is the $\eta$-causality condition is hold also. We will show this later on an example of a generic case.

Now, we will consider just a generic frame as well, when the final stage of the collapse is presented by the metric (\ref{generic}), in the Sect. \ref{G_generic}. To achieve this goal we use the transformation (\ref{EF_to_generic}) with arbitrary $f$ combined with (\ref{EF_to_PG}),
\be
cdt_p = cdt_f + \l(\frac{({r_g/r})^{1/2}}{1 + ({r_g/r})^{1/2}} - f({r_g}/{r})\r)dr= cdt_f + F({r_g}/{r})dr.
\m{PG_to_generic}
\ee
Then, for the extrinsic region $r>R(t)$ we obtain the results of the Sect. \ref{G_generic}. Namely, the metric (\ref{generic}) after the shift $t_f \goto t$ induce the field configuration (\ref{h-generic}) for that the non-zero at $r>R(t)$ total energy-momentum component (\ref{t-tot-00}) is constructed, where $f(r_g/r)$ satisfies the asymptotic behavior (\ref{first-f}).
To save a smooth matching between extrinsic and intrinsic regions one has to require that the function $F(r_g/r)$ in (\ref{PG_to_generic}) has to be differentiable and monotonic at $0< r \leq \infty$. Another requirement for the function $F(r_g/r)$ in (\ref{PG_to_generic}) is formulated as follows. After integrating (\ref{PG_to_generic}) and replacing $r$ by a surface radius $R(t_p)$ one can choose the constant of integration by the way that if the surface $R(t_p)$ goes to zero monotonically  at $t_p \goto 0$ then $R(t_f) \goto 0$ when $t_f \goto t \goto 0$. Then the final state (\ref{t-tot-00})-(\ref{t-tot-AB}) is achieved at $t \goto 0$. After satisfying these requirements, the model of the continuous collapse (\ref{PG}) plus (\ref{PG_in}) presented in the PG frame we rewrite in the generic frame with the use of the transformations (\ref{PG_to_generic}). One easily finds that after above discussion the requirements (i)-(iii) and (v) are hold again.

At last, let us turn to the $\eta$-causality problem. After transformation (\ref{PG_to_generic}) and the shift $t_f \goto t$ the metric (\ref{PG_in_tp}) for the intrinsic region acquires the form:
 \bea
 ds^2 = &-&\l(1- \frac{2m}{r}\r)dt^2 + 2 \l[\sqrt{\frac{2m}{r}} - \l(1- {\frac{2m}{r}}\r)F \r] dr dt\nonumber\\ &+& \l[1+2\sqrt{\frac{2m}{r}}F - \l(1- \frac{2m}{r}\r)F^2\r]dr^2 + r^2d\Omega^2\,.
\m{PG_in_tf}
 \eea
 For the sake of simplicity in formulae, here and below, we set $G=c=1$.
The standard exercises give the the expression for the ingoing ray of the light cone
\be
\l.\frac{dt}{dr}\r|_1=  -\frac{1}{1+ \sqrt{2m/r}} - F(r_g/r),
\m{cone_minus}
\ee
whereas the outgoing ray is determined by
\be
\l.\frac{dt}{dr}\r|_2=  \frac{1}{1- \sqrt{2m/r}} - F(r_g/r).
\m{cone_plus}
\ee
The necessary requirement for (\ref{cone_minus}) is that it has to be negative in all the regions. Then the $\eta$-causality condition, $\l.{dt}/{dr}\r|_1 \leq -1$, implies the restriction
\be
F \geq \frac{\sqrt{2m/r}}{1+ \sqrt{2m/r}}.
\m{cone_minus_eta}
\ee

To study (\ref{cone_plus}) it is necessary to consider three cases each of which corresponds to a concrete instant $t_p$.

1) The first case corresponds to the PG instant time when the star boundary $R(t_p) > r_g \equiv 2M/r$. Because $m\leq M$, and due to that $m$ in  (\ref{m_PG_ct}) is lowered with $r\goto 0$ at the instant $t_p$, one finds that
\be
\frac{2m}{r} <1.
\m{m<1}
\ee
Next, to be matched with the exterior smoothly the outgoing expression (\ref{cone_plus}) has to be positive. Then the $\eta$-causality condition, $\l.{dt}/{dr}\r|_2 \geq 1$, implies the restriction
\be
  \frac{\sqrt{2m/r}}{1- \sqrt{2m/r}} - F \geq 1.
\m{cone_plus_eta}
\ee
Combination of (\ref{cone_minus_eta}) and (\ref{cone_plus_eta}) gives the united restriction on $F(r_g/r)$
\be
\frac{\sqrt{2m/r}}{1+ \sqrt{2m/r}}\leq F \leq \frac{\sqrt{2m/r}}{1- \sqrt{2m/r}},
\m{cone_minus_plus_eta}
\ee
which is non-contradictable due to  (\ref{m<1}).

2) The second case is classified by the position of the star surface at the horizon $R(t_p) = r_g$.  Then (\ref{cone_plus}) gives $\l.{dt}/{dr}\r|_2 = + \infty$. It has the continuous matching with the extrinsic region, see (\ref{generic-cone2}). For the intrinsic region, where again the condition (\ref{m<1}) takes on the place, the result (\ref{cone_minus_plus_eta}) of the first case is repeated.

3) The third case that is classified by the position of the star surface, $R(t_p) < r_g$, is more complicated. The intrinsic region is decomposed into the three subregions: a) $2m/r>1$, b) $2m/r = 1$ and c) $2m/r <1$. In the case a) for the outgoing ray defined by (\ref{cone_plus}) one has $\l.{dt}/{dr}\r|_2 < 0$. Then, because the light cone must not to be degenerated, $\l.{dt}/{dr}\r|_1 > \l.{dt}/{dr}\r|_2$, one obtains
\be
- \frac{1}{1+ \sqrt{2m/r}}> \frac{1}{1- \sqrt{2m/r}}
\m{minus}
\ee
that is hold for the restriction of the case a) automatically. Analyzing subregions b) and c), one finds easily that the results correspond exactly to the results of the cases 2) and 1), respectively. Thus one obtains again the restriction (\ref{cone_minus_plus_eta}) only. Returning to the EF frame, one finds easily that for $F(r_g/r) = f_p(r_g/r)$ in (\ref{EF_to_PG_f}) the requirement (iv) discussed above is fulfilled.

To finalize the description of the continuous collapse in the field-theoretical formulation it is instructive to analyze the matter part $t^m_{\mu\nu}$ in  (\ref{t-T}) of the total energy-momentum. Thus, applying the transformations (\ref{PG_to_generic}) to the metric, $g^p_{\mu\nu}$, presented by (\ref{PG_in_tp}), we obtain $g^f_{\mu\nu}$ in (\ref{PG_in_tf}); applying the transformations (\ref{PG_to_generic}) to the energy-momentum, $T^p_{\mu\nu}$, presented in (\ref{T_00}), we obtain $T^f_{\mu\nu}$. Then, the formula (\ref{t-T}) acquires the form:
 \be
t^{m(f)}_{\mu\nu} = T^f_{\mu\nu} - \half
g^f_{\mu\nu}T^f - \half
\gamma_{\mu\nu}\gamma^{\alpha\beta}\l(T^f_{\alpha\beta} - \half
g^f_{\alpha\beta}T^f\r)\, .
 \m{t-T-f}
 \ee
 By the above consideration we conclude that, indeed, the intrinsic region defined by the energy-momentum (\ref{t-T-f}) is contracted at $t \goto 0$ to a point-like state described by the $\del$-function and presented in Sect. \ref{G_generic}.

 \section{Concluding remarks}
\m{discussion}

In the present paper, we develop the unique and complex strategy to represent the Schwarzschild black hole solution as a point mass particle including a continuous collapse to such an object. The field-theoretical approach is the basic technique in our study. Keeping in mind that the Schwarzschild black hole is an independent physical reality, we give another its description with making the use of an alternative equivalent mathematical language. We give  a generalized description restricted by physically reasonable requirements. This can be useful both from the fundamental point of view and for practical calculations.

We found out that the true singularity is not described by the $\delta$-distribution included into the energy density $t^{tot}_{00}$ {\em only}, like in many earlier approaches. The other components $t^{tot}_{11}$, see (\ref{t-tot-11}), and $t^{tot}_{AB}$, see (\ref{t-tot-AB}), could be interpreted as related to the ``intrinsic'' properties of the point, or ``intrinsic'' structure of the true singilarity. Indeed, they are proportional to $\delta({\bf r})$ as well,  and, thus, describe the point ``intrinsic radial'' and ``intrinsic tangent'' pressure.

To describe the Schwarzschild solution as a point particle we have used the exact equivalent of the Einstein equations in the form (\ref{FieldEqs}) without modification. This means that, basing on this simple point-like model, we could construct more complicated models and study them in the framework of the same (without, say, a regularization of various kinds) usual Einstein equations.


\begin{thebibliography}{99}

\bibitem{Landau_Lifshitz_1975} Landau, L.D., Lifshitz, E.M.: The Classical  Theory of Fields. Pergamon Press, Oxford (1975)

\bibitem{Misner_Thorn_Wheeler_1973} Misner, C.W., Thorne, K.S., Wheeler, J.A.: Gravitation. W.H. Freeman and Company, San Francisco (1973)

\bibitem{Balasin_Nachbagauer_1993} Balasin, H., Nachbagauer, H.: Class. Quantum Grav. {\bf 10}(11), 2271. Preprint arXiv::gr-qc/9305009 (1993)

\bibitem{Pantoja_Rago_2002} Pantoja, N.R., Rago, H.: Int. J. Mod. Phys. D {\bf 11}(9), 1479. Preprint arXiv:gr-qc/0009053 (2002)

\bibitem{Heinzle_Steinbauer_2002} Heinzle, J.M., Steinbauer, R.: J. Math. Phys. {\bf 43}(3), 1493. Preprint arXiv:gr-qc/0112047 (2002)

\bibitem{GW150914} Abbott, B.P., et al.: (LIGO-Virgo Scientific Collaborations), Phys. Rev. Lett. {\bf 116}(31 May 2016), 221101. Preprint: arXiv:1602.03841 [gr-qc] (2016)

\bibitem{GW151226} Abbott, B.P., et al.: (LIGO-Virgo Scientific Collaborations), Phys. Rev. Lett. {\bf 116}(15 June 2016), 241103. Preprint: arXiv:1606.04855 [gr-qc] (2016)

\bibitem{GW-A-LIGO} Abbott, B.P., et al.: (LIGO-Virgo Scientific Collaborations), Phys. Rev. X {\bf 6}(4), 241103. Preprint: arXiv:1606.04856 [gr-qc] (2016)

\bibitem{Damour_Jaranowski_Schafer_2001} Damour, T., Jaranowski, P., Sch\"afer, G.: Phys. Lett. B {\bf 513}(1-2), 147 (2001)

\bibitem{Schafer_2003} Sch\"afer, G.: Binary Black Holes and Gravitational Wave Production: Post-Newtonian Analytic Treatment. In: Fern\'{a}ndez-Jambrina, L., Gonz\'{a}lez-Romero, L.M. (eds.) Current Trends in Relativistic Astrophysics. Lecture Notes in Physics, vol. 617. Springer, Berlin (2003)

\bibitem{Schafer_2011} Sch\"afer, G.: Post-Newtonian Methods: Analytic Results on the Binary Problem. In: Blanchet, L., Spallicci, A., Whiting, B. (eds.) Mass and Motion in General Relativity. pp. 167-210. Springer, Dordrecht (2011)

\bibitem{Blanchet_2014}  Blanchet, L.: Living Rev. Relativ. {\bf 17}(2), 1. http://www.livingreviews.org/lrr-2014-2 (2014)

\bibitem{Narlikar_1985} Narlikar, J.V.: Some conceptual problems in general relativity and cosmology. In: Dadhich, N., Krishna R.J., Narlikar, J.V., Vishevara, C.V. (eds.) A Random Walk in Relativity and Cosmology. Viley Eastern Limited, New Delhi (1985)

\bibitem{GPP_1984} Grishchuk, L.P., Petrov, A.N., Popova, A.D.: Commun. Math. Phys. {\bf 94}(3), 379 (1984)

\bibitem{Popova_Petrov_1988} Popova, A.D., Petrov, A.N.: Int. J. Mod. Phys. A {\bf 3}(11), 2651 (1988)

\bibitem{Petrov_1993} Petrov, A. N.: Class. Quantum. Grav. {\bf 10}(12), 2663 (1993)

\bibitem{Petrov_2008} Petrov, A. N.: Nonlinear Perturbations and Conservation Laws on Curved Backgrounds in {G}{R} and Other Metric Theories. In: Christiansen, M.N., Rasmussen, T.K. (eds.) Classical and Quantum Gravity Research, pp. 79-160. Nova Science Publishers, New York. Preprint arXiv:0705.0019 [gr-qc] (2008)

\bibitem{Petrov+Co_2017} Petrov, A.N., Kopeikin, S.M., Lompay, R.R., Tekin, B.: Metric Theories of Gravity: Perturbations and Conservation Laws. De Gruyter, Germany (2017)

\bibitem{Geroch_Traschen_1987} Geroch, R., Traschen, J.: Phys. Rev. D {\bf 36}(15 August), 1017 (1987)

\bibitem{Colombeau_1983} Colombeau, J.: J. Math. Anal. Appl. {\bf 94}(1), 96 (1983)

\bibitem{Fiziev_2004} Fiziev, P.: The Gravitational Field of Massive Non-Charged Point Source in General Relativity. Preprint arXiv:gr-qc/0412131 (2004)

\bibitem{Goswami_Joshi_Vaz_Witten_2004} Goswami, R., Joshi, P.S., Vaz, C., Witten L.: Phys. Rev. D {\bf 70}(8), 084038 Preprint arXiv:gr-qc/0410041 (2004)

\bibitem{Cadoni_Magnemi_2005} Cadoni, M., Magnemi, S.: Mod. Phys. Lett. A {\bf 20}(38), 2919. Preprint arXiv:gr-qc/0503059 (2005)

\bibitem{Lundgren_Schmekel_York_2007} Lundgren, A.P., Schmekel, B.S., York, J.W.: Phys. Rev. D {\bf 75}(8), 084026. Preprint arXiv:gr-qc/0610088 (2007)

\bibitem{Brown_York_1993} Brown, J.D., York, J.W.: Phys. Rev. D {\bf 47}, 1407 (1993)

\bibitem{Katanaev_2013} Katanaev, M.O.: Gen. Relat. Grav. {\bf 45}(10), 1861. Preprint arXiv:1207.3481 [gr-qc] (2013)

\bibitem{Pitts_Schive_2004} Pitts, J.B., Schieve, W.C.: Found. Phys. {\bf 34}(2), 211. Preprint arXiv:gr-qc/0406102 (2004)

\bibitem{Petrov_Narlikar_1996} Petrov, A.N., Narlikar, J.V.: Found. Phys. {\bf 26}(9), 1201. (1996)

\bibitem{Petrov_2005} Petrov, A.N.: Found. Phys. Lett. {\bf 18}(5), 477. Preprint arXiv:gr-qc/0503082 (2005)

\bibitem{Deser_1970} Deser, S.: Gen. Relat. Grav. {\bf 1}(1), 9. Preprint arXiv:gr-qc/0411023 (1070)

\bibitem{Gelfand_Shilov_1958} Gelfand, I.M., Shilov, G.E.: Generalized functions. Vol. 1. Properties and Operations. Academic Press, New York (1964)

\bibitem{Bizon_Malec_1986} Bizon, P., Malec, E.: Class. Quantum Grav. {\bf 3}, L123 (1986)

\bibitem{OMurchadha_1986} O'Murchadha, N.: J. Math. Phys. {\bf 27}(8), 2111 (1986)

\bibitem{Kennefick_OM_1995} Kennefick, D. O'Murchadha, N.: Class. Quantum. Grav. {\bf 12}(1), 149 (1995)

\bibitem{Soloviev_1985} Soloviev, V. O.: Theor. Math. Phys. {\bf 65}, 1240 (1985)

\bibitem{Petrov_1995} Petrov, A. N.: Int. J. Mod. Phys. D {\bf 4}(4), 451 (1995)

\bibitem{Petrov_1997} Petrov, A. N.: Int. J. Mod. Phys. D {\bf 6}(2), 451 (1997)

\bibitem{Eddington_1924} Eddington, A. S.: Nature {\bf 113}(February 9), 192 (1924)

\bibitem{Finkelstein_1958} Finkelstein, D.: Phys. Rev {\bf 110}(4), 965 (1958)

\bibitem{Oppenheimer_Snyder_1939} Oppenheimer, J.R., Snyder, H.: Phys. Rev {\bf 56}(September 1), 455 (1939)

\bibitem{Kanai_Siino_Hosoya_2011} Kanai, Y., Siino, M., Hosoya, A.: Prog. Theor. Phys. {\bf 125}(May 1), 1053. Preprint arXiv:1008.0470 [gr-qc] (2011)

\bibitem{Peinleve_1921} Peinlev\'e, P.: C. R. Acad. Sci. (Paris) {\bf 173}(October 24), 677 (1921)

\bibitem{Gullstrand_1922} Gullstrand, A.: Arkiv. Mat. Astron. Fys. {\bf 16}(8), 1 (1922)

\bibitem{Hamilton_Lisle_2008} Hamilton, A.J.S., Lisle, J.P.: Am. J. Phys. {\bf 76}(6), 519 (2008)

\bibitem{Lin_Soo_2009} Lin, C.-Y., Soo, C.: Phys. Lett. B {\bf 671}(4-5), 493. Preprint arXiv:0810.2161 [gr-qc] (2009)

\bibitem{Jaén_Molina_2016} Jaén, X., Molina, A.: Rigid covariance as a natural extension of Painlevé--Gullstrand space-times: gravitational waves. Preprint arXiv:gr-qc/0411060 (2016)


\end{thebibliography}

\end{document}